\documentclass[aps,prl,showpacs,floatfix,twocolumn,byrevtex,superscriptaddress]{revtex4-1}
\usepackage{amsmath}
\usepackage{amssymb}
\usepackage{amstext}
\usepackage{amsopn}
\usepackage{amsfonts}
\usepackage{amsxtra}
\usepackage[english]{babel}
\usepackage{graphicx}
\usepackage{float}
\usepackage{bm}
\usepackage{multirow}
\usepackage{dcolumn}
\usepackage{color}
\usepackage{hyperref}

\newcommand{\icm}{\ensuremath{\textrm{cm}^{-1}}}

\newcommand{\LNO}{La$_{4}$Ni$_{3}$O$_{10}$}

\begin{document}

\title{Highly Anisotropic Charge Dynamics and Spectral Weight Redistribution in the Trilayer Nickelate La$_{4}$Ni$_{3}$O$_{10}$}
\author{Zhe~Liu}
\thanks{These authors contributed equally to this work.}
\author{Jie~Li}
\thanks{These authors contributed equally to this work.}
\affiliation{National Laboratory of Solid State Microstructures and Department of Physics, Collaborative Innovation Center of Advanced Microstructures, Nanjing University, Nanjing 210093, China}
\author{Deyuan~Hu}
\thanks{These authors contributed equally to this work.}
\affiliation{Center for Neutron Science and Technology, Guangdong Provincial Key Laboratory of Magnetoelectric Physics and Devices, School of Physics, Sun Yat-Sen University, Guangzhou, Guangdong 510275, China}
\author{Bingke~Ji}
\author{Haoran~Zhang}
\affiliation{National Laboratory of Solid State Microstructures and Department of Physics, Collaborative Innovation Center of Advanced Microstructures, Nanjing University, Nanjing 210093, China}
\author{Jiahao~Hao}
\affiliation{Key Laboratory for Microstructural Material Physics of Hebei Province, School of Science, Yanshan University, Qinhuangdao 066004, China}
\author{Yaomin~Dai}
\email{ymdai@nju.edu.cn}
\author{Qing~Li}
\author{Mengjun~Ou}
\affiliation{National Laboratory of Solid State Microstructures and Department of Physics, Collaborative Innovation Center of Advanced Microstructures, Nanjing University, Nanjing 210093, China}
\author{Bing~Xu}
\affiliation{Beijing National Laboratory for Condensed Matter Physics, Institute of Physics, Chinese Academy of Sciences, P.O. Box 603, Beijing 100190, China}
\author{Yi~Lu}
\email{yilu@nju.edu.cn}
\affiliation{National Laboratory of Solid State Microstructures and Department of Physics, Collaborative Innovation Center of Advanced Microstructures, Nanjing University, Nanjing 210093, China}
\author{Meng~Wang}
\email{wangmeng5@mail.sysu.edu.cn}
\affiliation{Center for Neutron Science and Technology, Guangdong Provincial Key Laboratory of Magnetoelectric Physics and Devices, School of Physics, Sun Yat-Sen University, Guangzhou, Guangdong 510275, China}
\author{Hai-Hu~Wen}
\email{hhwen@nju.edu.cn}
\affiliation{National Laboratory of Solid State Microstructures and Department of Physics, Collaborative Innovation Center of Advanced Microstructures, Nanjing University, Nanjing 210093, China}

\date{\today}
%
%

\begin{abstract}
We study the $ab$-plane and $c$-axis charge dynamics of La$_{4}$Ni$_{3}$O$_{10}$ using optical spectroscopy. While a pronounced Drude profile, i.e. metallic response, is observed in the $ab$-plane optical conductivity $\sigma_{1}^{ab}(\omega)$, the $c$-axis optical spectra $\sigma_{1}^{c}(\omega)$ exhibit semiconducting behavior. The zero-frequency extrapolation of the optical conductivity $\sigma_{1}(\omega \rightarrow 0) \equiv 1/\rho_{\text{dc}}$ gives a resistivity anisotropy of $\rho_{c}/\rho_{ab} \simeq 366$ at 300~K for La$_{4}$Ni$_{3}$O$_{10}$, which is much larger than the values in iron-based superconductors but comparable to those in high-$T_{c}$ cuprates. The interband response is also highly anisotropic, showing salient orbital selectivity for light polarized in the $ab$ plane and along the $c$ axis. The interband-transition peaks in both $\sigma_{1}^{ab}(\omega)$ and $\sigma_{1}^{c}(\omega)$ are located at lower energies compared to density-functional-theory predictions, signifying considerable electronic correlations. By investigating the spectral weight transfer, we find that in the pristine phase, Coulomb correlations have a marked impact on the charge dynamics of \LNO, whereas in the density-wave state, a gap opens with the Ni-$d_{z^{2}}$ orbital being involved.
\end{abstract}



\maketitle

%
%
The high-temperature (high-$T_{c}$) superconductivity in cuprates is achieved by doping a Mott insulator, i.e. the so-called parent compound which is characterized by Cu$^{2+}$ ions with a 3$d^{9}$ electron configuration coordinated in a square lattice of O atoms~\cite{Lee2006RMP,Imada1998RMP}. In light of this fact, Anisimov \emph{et al.}~\cite{Anisimov1999PRB} pointed out that nickelates with Ni$^{+}$ (3$d^{9}$) ions embedded in a planar O coordination constitute an analogue of the cuprate parent compounds, and doping these parent nickelates with low-spin Ni$^{2+}$ (3$d^{8}$) holes may lead to high-$T_{c}$ superconductivity. Along this line, superconductivity with a $T_{c}$ of approximately 15~K was realized in the thin films of hole-doped infinite-layer $R_{1-x}A_{x}$NiO$_{2}$ ($R$ = La, Pr, Nd; $A$ = Sr, Ca)~\cite{Li2019Nature,Li2020PRL,Zeng2020PRL,Osada2020NL,Osada2021AM,Sun2023AM,Zeng2022SA} and quintuple-layer square-planar Nd$_{6}$Ni$_{5}$O$_{12}$~\cite{Pan2022NM}.

Recently, the bilayer La$_{3}$Ni$_{2}$O$_{7}$~\cite{Sun2023Nature,Wang2024PRX,Zhang2024NP} was found to exhibit superconductivity with a $T_{c}$ up to 80~K under high pressure, igniting a new wave of investigations into nickel-based superconductors. Subsequently, superconductivity with a maximum $T_{c}$ of 30~K was also reported in the trilayer \LNO\ at 69~GPa~\cite{Zhu2024Nature,Zhang2025PRX,Li2024CPL,Li2024SCPMA}. At ambient pressure, both compounds are not superconducting but undergo density-wave transitions at low temperatures~\cite{Liu2023SCPMA,Zhang2020NC,Chen2024NC,Chen2024PRL,Zhao2025SB,Liu2024NC,Khasanov2025NP}. Unlike cuprates which are composed of square-planar CuO$_{2}$ planes, La$_{3}$Ni$_{2}$O$_{7}$ and \LNO\ belong to the Ruddlesden-Popper (RP) nickelate family La$_{n+1}$Ni$_{n}$O$_{3n+1}$, which consists of $n$ consecutive layers of corner-sharing NiO$_{6}$ octahedra alternating with rock salt LaO layers. In addition, La$_{3}$Ni$_{2}$O$_{7}$ and \LNO\ have electron configurations of 3$d^{7.5}$ and 3$d^{7.33}$, respectively, entirely distinct from the 3$d^{9}$ configuration in cuprates. While the active states in high-$T_{c}$ cuprates are dominated by the 3$d_{x^2-y^2}$ orbital~\cite{Lee2006RMP,Imada1998RMP}, La$_{3}$Ni$_{2}$O$_{7}$ and \LNO\ have active 3$d_{x^2-y^2}$ and 3$d_{z^2}$ orbitals near the Fermi level $E_{\text{F}}$~\cite{Sun2023Nature,Luo2023PRL,Lechermann2023PRB,Shilenko2023PRB,Zhang2023PRB,Tian2024PRB,Yang2024NC,Wang2024PRB,Li2017NC}. The question of whether RP nickelates represent a system analogous to high-$T_{c}$ cuprates is currently a subject of considerable debate, particularly regarding the dominant interactions, pairing mechanism and pairing symmetry~\cite{Yang2023PRBWang,Yang2023PRBZhang,Yang2024PRB,Jiang2024PRL,Fan2024PRBXiang,Zhang2024NC,Lu2024PRL}. In this context, a comprehensive investigation into the charge dynamics of the RP nickelates may shed light on the dominant correlations, as well as the differences and similarities between RP nickelates and cuprates.

\begin{figure*}[tb]
\includegraphics[width=\textwidth]{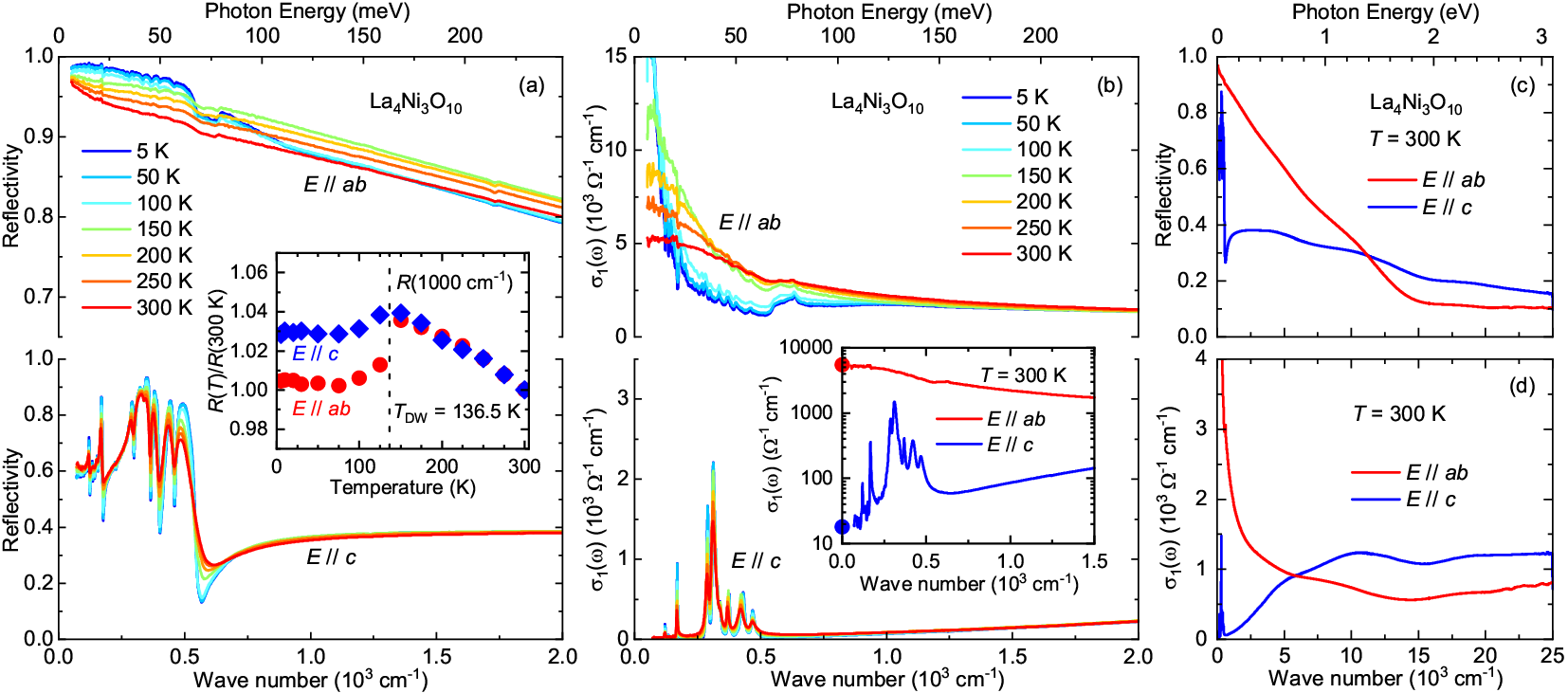}
\caption{(a) Reflectivity of \LNO\ at different temperatures for the $ab$ plane $R^{ab}(\omega)$ and $c$ axis $R^{c}(\omega)$. The inset shows the temperature dependence of $R^{ab}(\omega)$ (red solid circle) and $R^{c}(\omega)$ (blue solid diamond) at $\omega$ = 1000~\icm. (b) The $ab$-plane optical conductivity $\sigma^{ab}_{1}(\omega)$ and $c$-axis optical conductivity $\sigma^{c}_{1}(\omega)$ of \LNO\ at different temperatures. The inset displays $\sigma^{ab}_{1}(\omega)$ and $\sigma^{c}_{1}(\omega)$ at 300~K on the logarithmic scale with the zero-frequency extrapolations being denoted by solid circles. (c) and (d) show $R(\omega)$ and $\sigma_{1}(\omega)$ at 300~K along different directions in a broad frequency range up to 25\,000~\icm, respectively.}
\label{ARefS1}
\end{figure*}
We investigate the $ab$-plane and $c$-axis optical properties of \LNO. In the $ab$ plane, we found metallic behavior, as evidenced by a conspicuous Drude profile in the $ab$-plane optical conductivity $\sigma_{1}^{ab}(\omega)$. In sharp contrast, the absence of a Drude response in the $c$-axis optical conductivity $\sigma_{1}^{c}(\omega)$ indicates semiconducting behavior along the $c$ axis. Since $\sigma_{1}(\omega \rightarrow 0) \equiv 1/\rho_{\text{dc}}$, by extrapolating $\sigma_{1}^{ab}(\omega)$ and $\sigma_{1}^{c}(\omega)$ to zero frequency, we obtained $\rho_{c}/\rho_{ab} \simeq 366$ at 300~K. Such a large resistivity anisotropy is reminiscent of high-$T_{c}$ cuprates. Light polarized in the $ab$ plane and along the $c$ axis selectively excites interband transitions between different orbitals, giving rise to a strong anisotropy of the interband response. A detailed examination of the spectral weight redistribution in $\sigma_{1}^{ab}(\omega)$ and $\sigma_{1}^{c}(\omega)$ reveals that the charge dynamics in the pristine phase is strongly influenced by Coulomb correlations, whereas the density-wave state is characterized by the opening of a gap which involves the Ni-$d_{z^{2}}$ orbital.

%
%

%
%
Sample growth, characterization, optical measurements and data analysis methods are detailed in the Supplemental Materials~\cite{SuppMat}. Transport measurements indicate that the intertwined charge- and spin-density-wave transition occurs at $T_{\text{DW}}$ = 136.5~K in our \LNO\ sample (Supplemental Materials~\cite{SuppMat}). Figure~\ref{ARefS1}(a) shows the reflectivity of \LNO\ for light polarized in the $ab$ plane ($E \parallel ab$) $R^{ab}(\omega)$ and along the $c$ axis ($E \parallel c$) $R^{c}(\omega)$ up to 2000~\icm\ at several representative temperatures. $R^{ab}(\omega)$ is high and approaches unity in the low-frequency limit. Additionally, it increases as $T$ is lowered. All these properties suggest a metallic nature of \LNO\ in the $ab$ plane. Below $T_{\text{DW}}$, $R^{ab}(\omega)$ is significantly suppressed in a broad frequency range above 500~\icm\ due to the density-wave gap opening, in agreement with a previous optical study~\cite{Xu2025PRB}. By contrast, $R^{c}(\omega)$ attains a constant value much lower than 1 ($\sim$0.6) in the low-frequency limit and is dominated by a series of sharp features arising from IR-active phonon modes in the far-infrared range, which are prototypical characteristics of semiconducting materials~\cite{Jiang2020PRB,Hao2021PRB,Lobo2007PRB}. The inset of Fig.~\ref{ARefS1}(a) plots the $T$ dependence of $R^{ab}(\omega)$ (red solid circles) and $R^{c}(\omega)$ (blue solid diamonds) at $\omega$ = 1000~\icm. A drop is observed below $T_{\text{DW}}$ in both $R^{ab}(\omega)$ and $R^{c}(\omega)$, indicating that both the $ab$-plane and $c$-axis charge dynamics are affected by the density-wave transition in \LNO. Figure~\ref{ARefS1}(c) displays $R^{ab}(\omega)$ (red solid curve) and $R^{c}(\omega)$ (blue solid curve) at 300~K up to 25\,000~\icm, highlighting the strong anisotropy of \LNO\ in a broad frequency range.

\begin{figure*}[tb]
\includegraphics[width=\textwidth]{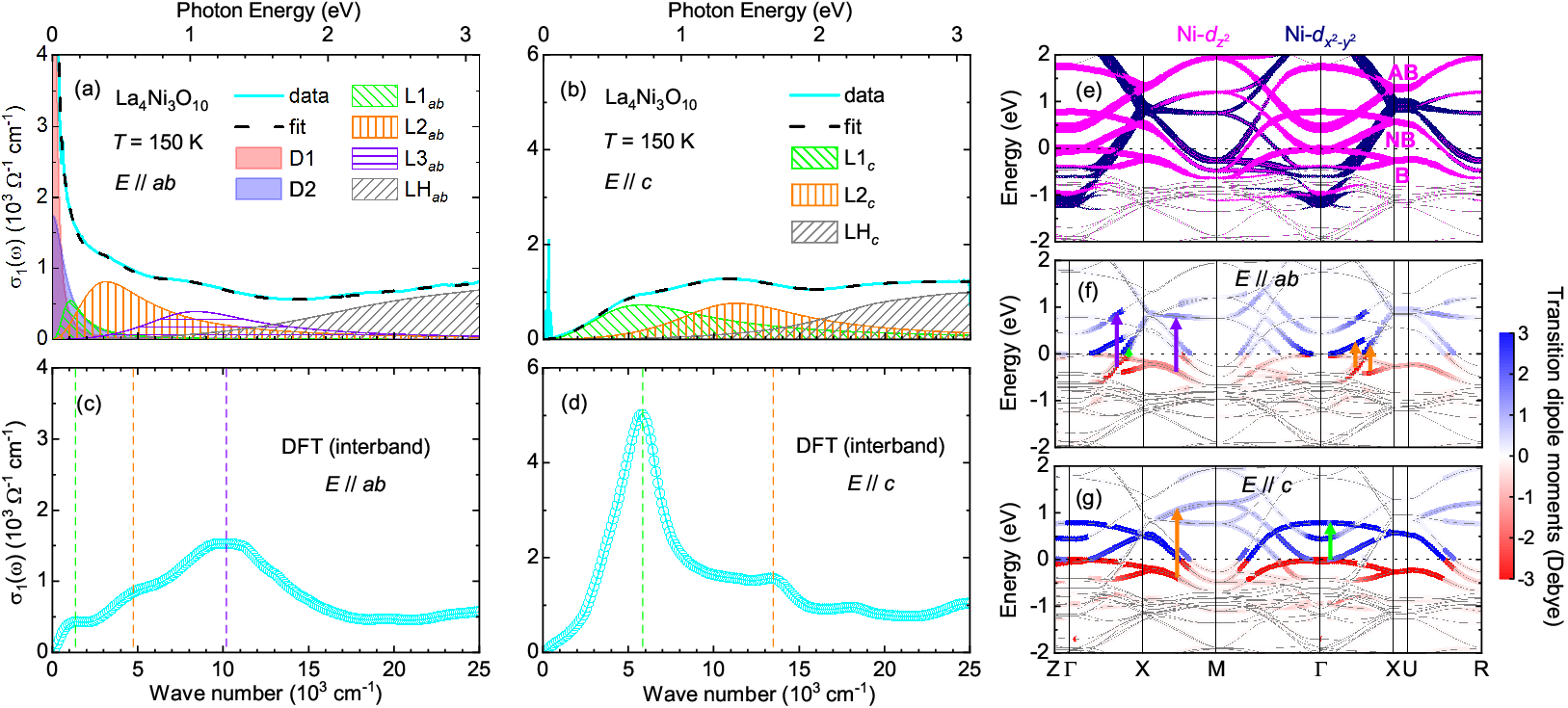}
\caption{(a) The measured $\sigma_{1}^{ab}(\omega)$ at 150~K (cyan solid curve) and the Drude-Lorentz fit (black dashed line). The fit is decomposed into two Drude components D1 (red shaded area), D2 (blue shaded area), and a series of Lorentz components L1$_{ab}$ (green hatched area), L2$_{ab}$ (orange hatched area), L3$_{ab}$ (violet hatched area) and LH$_{ab}$ (grey hatched area). (b) The measured $\sigma_{1}^{c}(\omega)$ at 150~K (cyan solid line) and the fit (black dashed line), which consists of several Lorentz components L1$_{c}$ (green hatched area), L2$_{c}$ (orange hatched area) and LH$_{c}$ (grey hatched area). (c) and (d) show the calculated interband $\sigma_{1}^{ab}(\omega)$ and $\sigma_{1}^{c}(\omega)$, respectively. The vertical dashed lines mark the positions of interband-transition peaks. (e) The calculated electronic band structure of \LNO, with magenta and navy denoting the Ni-$d_{z^{2}}$ and Ni-$d_{x^{2}-y^{2}}$ orbitals, respectively. Energy- and $\vec{k}$-resolved transition dipole moments for (f) $E \parallel ab$ and (g) $E \parallel c$. The initial and final states are colored with red and blue, respectively.}
\label{AFitCal}
\end{figure*}
%

Figure~\ref{ARefS1}(b) depicts the $ab$-plane optical conductivity $\sigma_{1}^{ab}(\omega)$ and $c$-axis optical conductivity $\sigma_{1}^{c}(\omega)$ of \LNO\ up to 2000~\icm\ at various temperatures. While the low-frequency $\sigma_{1}^{ab}(\omega)$ is characterized by a pronounced Drude response (a peak centered at zero frequency) which is a hallmark of metallic behavior, no Drude response is observed in $\sigma_{1}^{c}(\omega)$. Instead, the far-infrared $\sigma_{1}^{c}(\omega)$ is dominated by sharp phonon peaks, indicating a semiconducting nature of \LNO\ along the $c$ axis. Such semiconducting behavior in $\sigma_{1}^{c}(\omega)$ may be ascribed to the existence of insulating rock salt LaO layers, which effectively prevent the formation of coherence between different trilayers along the $c$ directions. The striking difference between $\sigma_{1}^{ab}(\omega)$ and $\sigma_{1}^{c}(\omega)$ points to highly anisotropic charge dynamics in \LNO. The zero-frequency extrapolation of $\sigma_{1}(\omega)$ yields the dc conductivity $\sigma_{1}$($\omega \rightarrow$ 0) $\equiv 1/\rho_{\textrm{dc}}$. The inset of Fig.~\ref{ARefS1}(b) plots $\sigma_{1}^{ab}(\omega)$ and $\sigma_{1}^{c}(\omega)$ at 300~K on a logarithmic scale; the red and blue solid circles on the vertical axis denote $\sigma_{1}^{ab}$($\omega\rightarrow$0) $\simeq$ 5490~$\Omega^{-1}\icm$ and $\sigma_{1}^{c}$($\omega\rightarrow$0) $\simeq$ 15~$\Omega^{-1}\icm$, respectively, which result in a resistivity anisotropy of $\rho_{c}$/$\rho_{ab}$ $\simeq$ 366 at 300~K. This value is significantly larger than those of iron-based superconductors (FeSCs), such as FeSe ($\sim$3--4)~\cite{Vedeneev2013PRB}, LiFeAs ($\sim$1--3)~\cite{Song2010APL}, and the BaFe$_{2}$As$_{2}$ family ($\sim$2--6)~\cite{Tanatar2009PRB,TanatarM2009PRB,Nakajima2018PRB}, but comparable to those of high-$T_{c}$ cuprates~\cite{Komiya2002PRB,Watanabe1997PRL}. Figure~\ref{ARefS1}(d) compares $\sigma^{ab}_{1}(\omega)$ (red solid curve) and $\sigma^{c}_{1}(\omega)$ (blue solid curve) over a broad frequency range at 300~K, revealing that the interband electronic transitions are also strongly anisotropic in \LNO.

We fit the experimental $\sigma_{1}(\omega)$ to the Drude-Lorentz model (see Supplemental Materials~\cite{SuppMat} and Refs.~\cite{Dressel2002,Tanner2019}). Figure~\ref{AFitCal}(a) displays the measured $\sigma_{1}^{ab}(\omega)$ at 150~K (cyan solid curve) and the fitting result (black dashed line), which is decomposed into two Drude components (D1 and D2) associated with intraband transitions, and a series of Lorentzian oscillators (L1$_{ab}$, L2$_{ab}$, L3$_{ab}$ and LH$_{ab}$) originating from interband transitions. In contrast, the fit of $\sigma_{1}^{c}(\omega)$ at 150~K [Fig.~\ref{AFitCal}(b)] only requires several Lorentz components (L1$_{c}$, L2$_{c}$ and LH$_{c}$) to describe the interband transitions, as Drude behavior is absent in the $\sigma_{1}^{c}(\omega)$ spectrum of \LNO.

In order to elucidate the origin of these components in $\sigma_{1}^{ab}(\omega)$ and $\sigma_{1}^{c}(\omega)$, we calculated the orbital-resolved electronic band structure of \LNO\ (Supplemental Materials~\cite{SuppMat}). As shown in Fig.~\ref{AFitCal}(e), the bands near $E_{\textrm{F}}$ (within $\pm$2~eV) are dominated by Ni-$d_{x^{2}-y^{2}}$ (navy) and Ni-$d_{z^{2}}$ (magenta) orbitals, similar to the bilayer La$_{3}$Ni$_{2}$O$_{7}$~\cite{Abadi2025PRL,Christiansson2023PRL,Geisler2024npjQM,LaBollita2023arXiv,Liu2024NC,LiY2024CPL,Luo2023PRL,Sun2023Nature,Yang2024NC,Zhang2023PRB}. Due to strong hybridization between the Ni-$d_{z^{2}}$ and O-$p_{z}$ orbitals, the bands originating from the Ni-$d_{z^{2}}$ orbital are split into anti-bonding (AB), non-bonding (NB) and bonding (B) branches~\cite{Li2024SCPMA,Sakakibara2024PRB,Chen2024PRB,Zhang2024PRLDagotto,Zhang2024PRB}, whereas the Ni-$d_{x^{2}-y^{2}}$ orbitals lying in the NiO$_{6}$ plane are only weakly coupled between planes, thus not undergoing the AB-NB-B splitting. While the multiple bands formed by the Ni-$d_{x^{2}-y^{2}}$ and Ni-$d_{z^{2}}$ orbitals crossing $E_{\textrm{F}}$ give rise to the two Drude components in $\sigma_{1}^{ab}(\omega)$, electronic transitions between these bands are responsible for the Lorentz components in $\sigma_{1}^{ab}(\omega)$ and $\sigma_{1}^{c}(\omega)$. To gain further insights into the interband transitions, we calculated the energy- and momentum-resolved transition dipole moments (TDMs) $P_{\alpha}^{2}(n, \vec{k})$ for $E \parallel ab$ [Fig.~\ref{AFitCal}(f)] and $E \parallel c$ [Fig.~\ref{AFitCal}(g)] (Supplemental Materials~\cite{SuppMat} and Refs.~\cite{Geisler2024npjQM}). A combination of the TDMs and the orbital-resolved band structure [Fig.~\ref{AFitCal}(e)] allows us to clarify the origin of the Lorentz components in $\sigma_{1}^{ab}(\omega)$ and $\sigma_{1}^{c}(\omega)$. For $E \parallel ab$, the interband transitions mainly occur between the Ni-$d_{x^{2}-y^{2}}$ and Ni-$d_{z^{2}}$ orbitals: L1$_{ab}$ stems from the transitions between the occupied Ni-$d_{z^{2}}$ bonding states and the empty Ni-$d_{x^{2}-y^{2}}$ bands (green arrow); L2$_{ab}$ arises from transitions from the occupied Ni-$d_{x^{2}-y^{2}}$ bands to the empty Ni-$d_{z^{2}}$ non-bonding bands and from the occupied Ni-$d_{z^{2}}$ bonding bands to the empty Ni-$d_{x^{2}-y^{2}}$ bands (orange arrows); the transitions from the occupied Ni-$d_{x^{2}-y^{2}}$ bands to the empty Ni-$d_{z^{2}}$ anti-bonding bands and from the occupied Ni-$d_{z^{2}}$ bonding bands to the empty Ni-$d_{x^{2}-y^{2}}$ bands (violet arrows) give rise to L3$_{ab}$. For $E \parallel c$, the interband excitations are dominated by electronic transitions between Ni-$d_{z^{2}}$ orbitals: both L1$_{c}$ and L2$_{c}$ originate from electronic transitions between the occupied Ni-$d_{z^{2}}$ bonding and the empty Ni-$d_{z^{2}}$ non-bonding bands, as denoted by the green and orange arrows in Fig.~\ref{AFitCal}(g). Both LH$_{ab}$ and LH$_{c}$ correspond to high-energy transitions involving bands far from $E_{\textrm{F}}$.

Figures~\ref{AFitCal}(c) and \ref{AFitCal}(d) show the calculated interband $\sigma_{1}(\omega)$ for $E \parallel ab$ and $E \parallel c$, respectively~\cite{SuppMat}. While the calculated $\sigma_{1}(\omega)$ qualitatively captures the interband components in the measured $\sigma_{1}(\omega)$ for both polarizations, there are noticeable discrepancies between the calculated and experimental $\sigma_{1}(\omega)$. Such discrepancies arise from the effects of electronic correlations that are not taken into account in the DFT calculations. In correlated materials, electronic correlations lead to a renormalization of the electronic bands, i.e. a reduction of the band width, which induces a shift of the interband transition peaks in the experimental $\sigma_{1}(\omega)$ towards lower energy compared to DFT calculations~\cite{Qazilbash2009NP,Si2009NP,Xu2020NC,Cao2025PRB,Yi2025PRL}. Here in \LNO, the Lorentz components in the measured $\sigma_{1}^{ab}(\omega)$ [Fig.~\ref{AFitCal}(a)] and $\sigma_{1}^{c}(\omega)$ [Fig.~\ref{AFitCal}(b)] all exhibit a noticeable redshift compared to the interband transition peaks in the calculated $\sigma_{1}^{ab}(\omega)$ [Fig.~\ref{AFitCal}(c)] and $\sigma_{1}^{c}(\omega)$ [Fig.~\ref{AFitCal}(d)], indicating that the bands formed by Ni-$d_{x^{2}-y^{2}}$ and Ni-$d_{z^{2}}$ orbitals are renormalized due to electronic correlation effects. It is also noteworthy that the redshift for each interband component is different from others, implying that the electronic-correlation-induced renormalization varies from band to band in \LNO. Furthermore, L3$_{ab}$ and L1$_{c}$ in the measured $\sigma_{1}(\omega)$ are significantly smaller or broader than the corresponding interband transition peaks in the calculated $\sigma_{1}(\omega)$, whereas other components agree relatively better with theoretical calculations. This fact also points to band-dependent electronic correlation effects. ARPES measurements~\cite{Li2017NC,Du2024arXiv} and DFT+DMFT calculations~\cite{Wang2024PRB} have found that the Ni-$d_{z^{2}}$ orbital features stronger electronic correlations than the Ni-$d_{x^{2}-y^{2}}$ orbital, compatible with our optical results.

\begin{figure}[tb]
\includegraphics[width=\columnwidth]{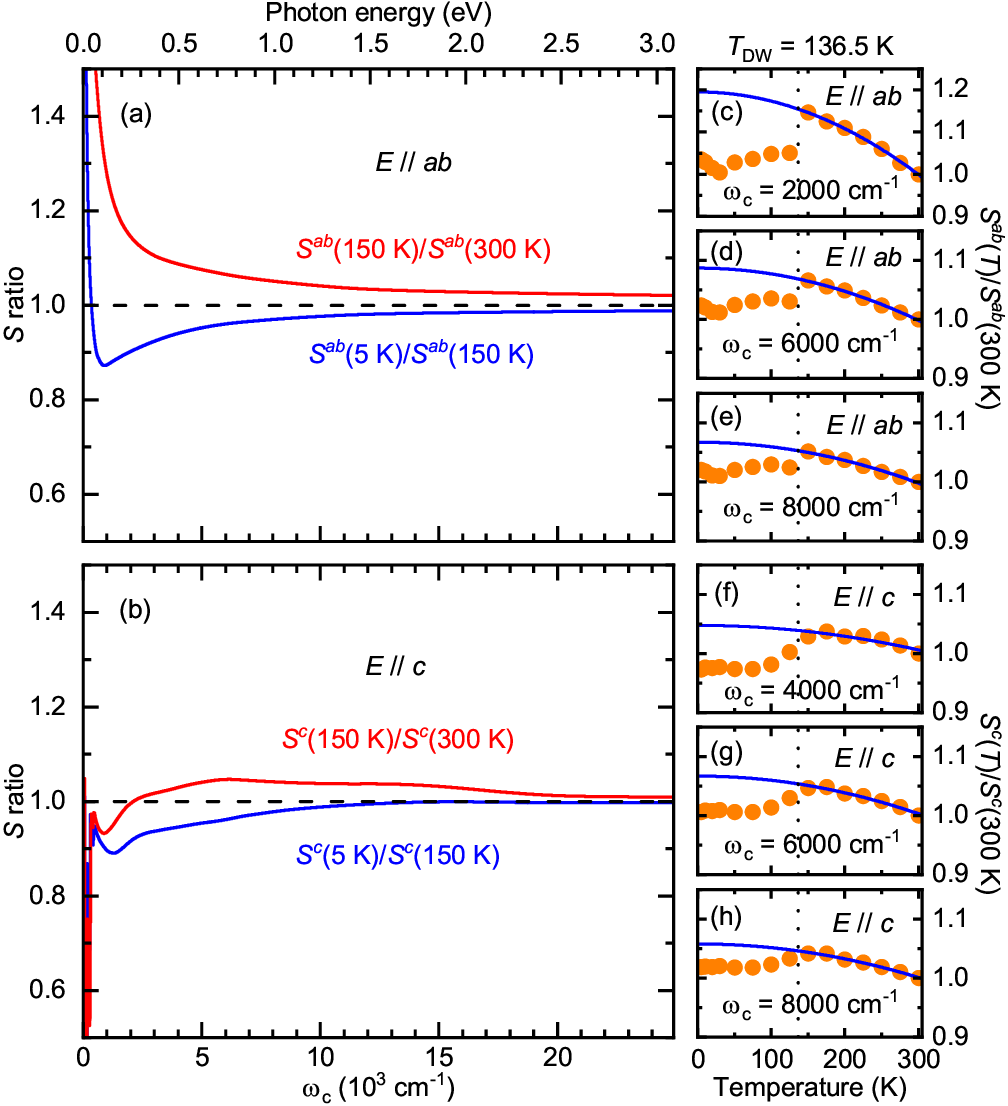}
\caption{The $S$ ratios between different temperatures as a function of $\omega_{c}$ for (a) $E \parallel ab$ and (b) $E \parallel c$. (c)-(h) The $T$ dependence of $S$ (normalized by its value at 300~K) for different cutoff frequencies.}
\label{AIntT}
\end{figure}

Next, we investigate the $T$ dependence of the charge dynamics by tracking the transfer of the spectral weight defined as
%
\begin{equation}
S = \int_{0}^{\omega_c}\sigma_{1}(\omega)d\omega,
\label{SW}
\end{equation}
where $\omega_{c}$ is a cutoff frequency. Figures~\ref{AIntT}(a) and \ref{AIntT}(b) display the $S$ ratios between different temperatures for $E \parallel ab$ and $E \parallel c$, respectively, which conveniently reflect the direction and energy scale of spectral weight transfer. Specifically, an $S$ ratio exceeding 1 corresponds to a spectral weight transfer from high to low energy, whereas a spectral weight transfer from low to high energy leads to an $S$ ratio lower than 1. When $\omega_{c}$ covers the full energy scale of the spectral weight transfer, the $S$ ratio reaches 1 due to the conservation of spectral weight.

$S^{ab}(150~\text{K})/S^{ab}(300~\text{K})$ is larger than 1 and approaches 1 with increasing $\omega_{c}$. Although $S^{c}(150~\text{K})/S^{c}(300~\text{K})$ is below 1 in the low-frequency range, which is most likely associated with the reduction of thermally excited carriers with decreasing $T$, it exceeds 1 for $\omega_{c} > 2000$~\icm. These observations suggest that in the pristine phase, as $T$ is reduced, there is a spectral weight transfer from high to low energy over a broad frequency range in both $\sigma_{1}^{ab}(\omega)$ and $\sigma_{1}^{c}(\omega)$. This effect can be further confirmed by the $T$ dependence of $S^{ab}$ and $S^{c}$: as shown in Figs.~\ref{AIntT}(c)-\ref{AIntT}(h), for different $\omega_{c}$'s, both $S^{ab}(T)/S^{ab}(300~\text{K})$ and $S^{c}(T)/S^{c}(300~\text{K})$ increase monotonically as $T$ decreases from 300~K to 150~K. In a Hund's metal, the decrease of $T$ results in a spectral weight transfer from low to high frequency. Such behavior has been reported in iron-based superconductors and attributed to the strong Hund's coupling between itinerant electrons and localized electron moments~\cite{Schafgans2012PRL,Wang2012JPCM,Wu2011PRB,Xu2014CPB}. On the contrary, Mott-Hubbard systems are expected to feature a spectral weight transfer from high to low frequency as $T$ is lowered~\cite{Kotliar2004PT,Toschi2005PRL}, which has been widely observed in cuprates~\cite{Molegraaf2002Science,Toschi2005PRL,Ortolani2005PRL,Carbone2006PRB1,Carbone2006PRB2,Santander-Syro2002PRL}, V$_{2}$O$_{3}$ in the high-temperature metallic state~\cite{Qazilbash2008PRB}, and Nd$_{4}$Ni$_{3}$O$_{8}$~\cite{Hao2021PRB}. The spectral weight redistribution in  \LNO\ hints that Coulomb correlations exert considerable influence on the charge dynamics which place \LNO\ closer to a Mott system rather than a Hund's system. It is also worth noting that in the pristine phase, the $T$ dependence of $S^{ab}$ and $S^{c}$ for all different $\omega_{c}$'s follows $T^{2}$ behavior [blue solid lines in Figs.~\ref{AIntT}(c)-\ref{AIntT}(h)], in accord with the calculations based on the dynamical mean-field theory of the Hubbard model~\cite{Toschi2005PRL} and the experimental results in cuprates~\cite{Molegraaf2002Science,Ortolani2005PRL,Carbone2006PRB1,Carbone2006PRB1,Hao2021PRB} and Nd$_{4}$Ni$_{3}$O$_{8}$~\cite{Hao2021PRB}.

While $S^{ab}(5~\text{K})/S^{ab}(150~\text{K})$ exceeds 1 in the very low-frequency range, which is due to the narrowing of the Drude response, it falls below 1 for $\omega_{c} > 500$~\icm; $S^{c}(5~\text{K})/S^{c}(150~\text{K})$ lies below 1 and approaches 1 as $\omega_{c}$ grows. These observations unambiguously suggest that for both $E \parallel ab$ and $E \parallel c$, a spectral weight transfer from low to high energy occurs at low temperatures. Furthermore, Figs.~\ref{AIntT}(c)-\ref{AIntT}(h) show that for different $\omega_{c}$'s, both $S^{ab}(T)/S^{ab}(300~\text{K})$ and $S^{c}(T)/S^{c}(300~\text{K})$ exhibit a drop below the density-wave transition at $T_{\text{DW}}$, attesting to the suppression of low-frequency spectral weight. All these results are in agreement with the optical response of a density-wave gap, which suppresses the low-frequency spectral weight, i.e. the Drude weight in $\sigma_{1}(\omega)$, and transfers it to higher frequency~\cite{Hu2008PRL,Zhou2021PRB,Liu2024NC,Xu2025PRB}. Here, it is worth noting that for $E \parallel c$, Drude behavior is absent, suggesting that the spectral weight transfer in $\sigma_{1}^{c}(\omega)$ originates from a redistribution of interband transitions. Since $\sigma_{1}^{c}(\omega)$ is dominated by interband transitions between the occupied Ni-$d_{z^2}$ bonding states and the empty Ni-$d_{z^2}$ non-bonding states [Fig.~\ref{AFitCal}(g)], the observation of density-wave-induced spectral weight redistribution in $\sigma_{1}^{c}(\omega)$ implies that the Ni-$d_{z^2}$ orbital is involved in the formation of the density-wave gap, highlighting the important role of the Ni-$d_{z^2}$ orbital in driving the density-wave instabilities. Note that previous ARPES studies on \LNO\ revealed contradictory results regarding the question of which bands are gapped by the density-wave transition~\cite{Li2017NC,Du2024arXiv}, calling for further investigations into the momentum dependence of the density-wave gap.

%
%

%
%
In summary, the $ab$-plane and $c$-axis charge dynamics of \LNO\ have been studied using optical spectroscopy. While $\sigma_{1}^{ab}(\omega)$ exhibits a pronounced Drude peak, indicating a metallic nature of \LNO\ in the $ab$ plane, semiconducting behavior is observed in $\sigma_{1}^{c}(\omega)$. The zero-frequency extrapolation of $\sigma_{1}^{ab}(\omega)$ and $\sigma_{1}^{c}(\omega)$ at 300~K yields a resistivity anisotropy of $\rho_{c}/\rho_{ab} \simeq 366$ for \LNO, which is much higher than the values in FeSCs, but similar to those in cuprates. The interband response also features strong anisotropy, as light polarized in the $ab$ plane and along the $c$ axis selectively excites electronic transitions between different orbitals. In the pristine phase, the spectral weight transfer implies that the charge dynamics is under the influence of Coulomb correlations, whereas in the density-wave state, the spectral weight redistribution suggests the opening of a gap with the participation of the Ni-$d_{z^{2}}$ orbital.

%
%

\begin{acknowledgments}
We thank Ilya M. Eremin, Frank Lechermann, Qianghua Wang and Shunli Yu for helpful discussions. Work at NJU was supported by the National Key R\&D Program of China (Grants No. 2022YFA1403201 and 2022YFA1403000), the National Natural Science Foundation of China (Grants No. 12174180 and 12574144). Work at SYSU was supported by the National Natural Science Foundation of China (Grant No. 12425404), the National Key Research and Development Program of China (Grant No. 2023YFA1406500), the CAS Superconducting Research Project (Grant No. SCZX-0101), the Guangdong Provincial Key Laboratory of Magnetoelectric Physics and Devices (Grant No. 2022B1212010008), and Research Center for Magnetoelectric Physics of Guangdong Province (Grant No. 2024B0303390001).
\end{acknowledgments}

%
%

\end{document}